\documentclass[aps,prd]{revtex4}

\usepackage{graphicx}
\usepackage{array}

\def\lg{{\mathchoice{~\raise.58ex\hbox{$<$}\mkern-14.8mu\lower.52ex\hbox{$>$}~}
                    {~\raise.58ex\hbox{$<$}\mkern-14.8mu\lower.52ex\hbox{$>$}~}
                    {\raise.59ex\hbox{{$\scriptscriptstyle <$}}\mkern-12.8mu%
                     \lower.01ex\hbox{{$\scriptscriptstyle >$}}}   {}   }}
\def\gl{{\mathchoice{~\raise.58ex\hbox{$>$}\mkern-12.8mu\lower.52ex\hbox{$<$}~}
                    {~\raise.58ex\hbox{$>$}\mkern-12.8mu\lower.52ex\hbox{$<$}~}
                    {\raise.62ex\hbox{{$\scriptscriptstyle >$}}\mkern-12.0mu%
                     \lower.05ex\hbox{{$\scriptscriptstyle <$}}}  {}    }}

\newcommand{\be}{\begin{equation}}
\newcommand{\ee}{\end{equation}}
\newcommand{\ba}{\begin{eqnarray}}
\newcommand{\ea}{\end{eqnarray}}
\newcommand{\ban}{\begin{eqnarray*}}
\newcommand{\ean}{\end{eqnarray*}}
\newcommand \nn {\nonumber}
\newcommand{\sla}{\!\!\!/ \,}

\begin{document}

\title{Collisional Processes in Supersymmetric Plasma}

\author{Alina Czajka}

\affiliation{Institute of Physics, Jan Kochanowski University,
25-406 Kielce, Poland}

\author{Stanis\l aw Mr\' owczy\' nski}

\affiliation{Institute of Physics, Jan Kochanowski University,
25-406 Kielce, Poland}
\affiliation{National Centre for Nuclear Research, 00-681 Warsaw, Poland}

\date{October 15, 2011}

\begin{abstract}

Collisional processes in ultrarelativistic ${\cal N} =1$ SUSY QED  plasma are studied and
compared to those in an electromagnetic plasma of electrons, positrons and photons.
Cross sections of all binary interactions which occur in the supersymmetric plasma at the 
order of $e^4$ are computed. Some processes, in particular the Compton scattering on 
selectrons, appear to be independent of momentum transfer and thus they are qualitatively 
different from processes in an electromagnetic plasma. It suggests that transport properties 
of the SUSY plasma are different than those of its non-supersymmetric counterpart. Energy loss 
and momentum broadening of a particle traversing the supersymmetric plasma are discussed 
in detail and the characteristics are shown to be surprisingly similar to those of QED plasma.

\end{abstract}



\maketitle

\section{Introduction}

This is our second paper where the ultrarelativistic ${\cal N} =1$ SUSY QED  plasma 
is studied and compared to an electromagnetic plasma of electrons, positrons and photons. 
In the first one \cite{Czajka:2010zh} we analyzed collective excitations of a supersymmetric 
plasma finding a rather surprising similarity between the two systems. Here we focus on 
collisional processes which control transport properties of the plasmas. 

Our motivation to study the supersymmetric plasma is twofold. First of all supersymmetry 
is a good candidate to be a symmetry of Nature at sufficiently high energies and if true,  
supersymmetric plasmas existed in the early Universe. Experiments at the Large Hadron 
Collider can soon show whether there is any evidence to support the idea of supersymmetry. 
However, independently of its ontological status, supersymmetric field theories are worth 
studying because of their unique features. The AdS/CFT duality of the five-dimensional 
gravity in the anti de Sitter geometry and the conformal field theories, see the review 
\cite{Aharony:1999ti}, revived a great interest in the ${\cal N} = 4$ supersymmetric 
Yang-Mills theory which is both classically and quantum mechanically conformally invariant. 

AdS/CFT duality has provided a method to study strongly coupled field theories and numerous 
interesting results have been obtained, see the reviews \cite{Son:2007vk,Janik:2010we}, but 
the relevance of the results for non-supersymmetric theories, which are of actual interest, 
is somewhat unclear. A systematic comparative analyses of supersymmetric systems and
their non-supersymmetric counterparts can be done in the domain of weak coupling where 
perturbative methods are applicable and several studies have been performed 
\cite{CaronHuot:2006te,Huot:2006ys,CaronHuot:2008uh,Blaizot:2006tk,Chesler:2006gr,Chesler:2009yg}. 
We are particularly interested in non-equilibrium plasmas, which have not attracted much
attention yet. We have started with the supersymmetric  ${\cal N} =1$ QED plasma 
which is noticeably simpler than that of ${\cal N} =4$ Super Yang-Mills.

As already mentioned, we discuss here collisional processes in the ${\cal N} =1$ QED plasma 
which is assumed to be ultrarelativistic and thus all particles are treated as massless. We first 
compute cross sections of all binary processes which occur at the lowest nontrivial order of 
$\alpha \equiv e^2/4\pi$. There appears to be a class of binary interactions in the SUSY 
plasma which are qualitatively different than those in the electromagnetic plasma of electrons, 
positrons and photons where the interactions with small momentum transfer are dominant. 
For example, Compton scattering on selectrons is isotropic in the center-of-mass frame 
while the usual Compton scattering on electrons mostly occurs at small angles. Since collisional 
processes determine transport properties of a many-body system, the supersymmetric plasma 
can be expected to be rather different than its non-supersymmetric counterpart. It should be 
remembered, however, that the temperature is the only dimensional parameter which 
characterizes the equilibrium ultrarelativistic plasma. Consequently, the parametric form 
of transport coefficients can be determined by dimensional arguments. For example, the 
shear viscosity must be proportional to $T^3/\alpha^2$ and it is thus hard to expect that the
viscosity of supersymmetric plasma is qualitatively different than that of electromagnetic 
one. Indeed, the shear viscosity of an ${\cal N} =4$ Super Yang Mills plasma is rather 
similar that of a quark-gluon plasma \cite{Huot:2006ys}. 

We consider here two transport characteristics of the ${\cal N} =1$ QED plasma which
are not so constrained by dimensional arguments. Specifically, we compute the collisional 
energy loss and momentum broadening of a particle traversing the equilibrium plasma. 
The latter quantity determines a magnitude of radiative energy loss of highly energetic particle
in a plasma \cite{Baier:1996sk}. The dimensional argument does not work here because the
two quantities depend not only on the plasma temperature but on the energy of test particle 
as well. We show that the energy loss and momentum broadening in SUSY plasma appear 
to be surprisingly similar to those in electromagnetic one. 

Our paper is organized as follows. In the subsequent section, we introduce ${\cal N}=1$ 
SUSY QED by writing down and discussing its lagrangian. In Sec.~\ref{sec-cross-binary} we 
enlist and analyze the binary processes which occur in the plasma.  Secs. \ref{sec-e-loss} 
and \ref{sec-qhat} are devoted to the problem of, respectively, energy loss and momentum 
broadening of a test particle. The paper is closed with a summary of our main results 
and conclusions.  The natural system of units with $c= \hbar = k_B =1$ and the signature 
of the metric tensor $(+ - - -)$ are used throughout the article.

\section{${\cal N}=1$ SUSY QED}
\label{sec-lagrangian}

The lagrangian of ${\cal N}=1$ SUSY QED is known, see {\it e.g.} \cite{Binoth:2002xg},
to be
\ba
\label{L-SUSY-QED}
{\cal L} &=& -\frac{1}{4}F^{\mu \nu} F_{\mu \nu} 
+  i\bar \Psi D\!\sla \Psi
+\frac{i}{2} \bar \Lambda \partial \sla \Lambda
+(D_\mu \phi_L)^*(D^\mu \phi_L) + (D_\mu^* \phi_R)(D^\mu \phi_R^*)
\\ \nn
&& +\sqrt{2} e \big( \bar \Psi P_R \Lambda \phi_L - \bar \Psi P_L \Lambda \phi_R^*
+ \phi_L^* \bar \Lambda P_L \Psi - \phi_R \bar \Lambda P_R \Psi \big)
- \frac{e^2}{2} \big( \phi_L^* \phi_L - \phi_R^* \phi_R \big)^2 ,
\ea
where the strength tensor $F^{\mu \nu}$ is expressed through the electromagnetic
four-potential $A^\mu$ as $F^{\mu \nu} \equiv \partial^\mu A^\nu - \partial^\nu A^\mu$
and the covariant derivative equals $D^\mu \equiv \partial^\mu +ie A^\mu$; $\Lambda$ is 
the Majorana bispinor photino field, $\Psi$ is the Dirac bispinor electron field, $\phi_L$ 
and $\phi_R$ are the scalar left selectron and right selectron fields; the projectors $P_L$ 
and $P_R$ are defined in a standard way $P_L \equiv \frac{1}{2}(1 - \gamma_5)$ and 
$P_R \equiv \frac{1}{2}(1 + \gamma_5)$. Since we are interested in ultrarelativistic 
plasmas, the mass terms are neglected in the lagrangian. 

As seen in the lagrangian (\ref{L-SUSY-QED}), there is a self-interaction of selectron field
due to the terms $ ( \phi_L^* \phi_L )^2$, $( \phi_R^* \phi_R)^2$ and 
$-2 \phi_L^* \phi_L \phi_R^* \phi_R$. There is also a four-boson coupling 
$ \phi_{L,R}^* \phi_{L,R} A^\mu A_\mu$ of the selectron with the electromagnetic
field. Such a contact interaction is qualitatively different  than that caused by a massless 
particle exchange. The scattering cross section in the absence of other interactions
is isotropic in the center-of-mass frame of colliding particles and the energy and 
momentum transfers are bigger than that in electromagnetic interactions. Therefore, one 
expects that consequently transport properties of supersymmetric ${\cal N} =1$ QED plasma differ 
from those of  QED plasma of electrons, positrons and photons. To test this expectation, 
in the next section we compute the cross sections of binary processes which occur in 
${\cal N} =1$ SUSY QED  plasma in the lowest non-trivial order of the coupling constant 
$\alpha \equiv e^2/4\pi$. Further on, the cross sections are used to derive the energy
loss and momentum broadening of an energetic  particle traversing the plasma.

\section{Cross sections of binary interactions}
\label{sec-cross-binary}

We discuss here the cross sections of all binary interactions contributing at the order of 
$\alpha^2$ ($\alpha \equiv e^2/4\pi$). The processes along with corresponding Feynman diagrams 
and cross sections are listed in Table~\ref{tab-corr-binray}. Electrons, selectrons,
photons and photinos are denoted as $e$, $\tilde{e}$, $\gamma$, $\tilde{\gamma}$. 
In every process the initial and/or final state particles can carry positive or negative 
charge; selectrons can be additionally `$R$' (right) or `$L$' (left). For each listed 
reaction the Feynman diagrams from the third column correspond to the first combination 
of charges of interacting particles. For example, the Feynman diagrams of the process 
$e^{\mp}e^{\mp} \longrightarrow e^{\mp}e^{\mp}$ actually represent the scattering 
$e^- e^- \longrightarrow e^- e^-$. The solid, dashed, wavy and double-solid lines in 
the Feynman diagrams correspond to electrons, selectrons, photons and photinos, respectively.  

The computed cross sections, which are averaged over initial polarizations of colliding 
particles and summed over polarizations of final state particles, are expressed through 
the Mandelstam invariants $s,t$ and $u$ defined in the standard way. For a process denoted 
as $ab \longrightarrow cd$, we have
\be
s \equiv (p+p_1)^2 , \;\;\;\;\;\;  t \equiv (p - p')^2 , \;\;\;\;\;\;  u \equiv (p - p'_1)^2,
\ee
where $p, p_1, p', p'_1$ are four-momenta of particles $a, b, c, d$, respectively.
The corresponding Feynman diagrams are drawn in such a way that particle $a$ comes 
from the upper left corner of the diagram, particle $b$ comes from the lower left 
corner, particle $c$ goes to the upper right, and  particle $d$ goes to the 
lower right corner. 

The first five processes from Table~\ref{tab-corr-binray} occur in a supersymmetric 
QED plasma and in an electromagnetic plasma of electrons, positrons and photons. The 
remaining process are characteristic for the  ${\cal N} =1$ SUSY QED  plasma. As can be 
seen from Table~\ref{tab-corr-binray}, the processes No. 6 -- 8 and 23 -- 26 are independent of 
momentum transfer. As a result, the scattering is isotropic in the center-of-mass frame of colliding 
particles. Such processes are qualitatively different from those in an electromagnetic 
plasma (processes No. 1 -- 5) which are dominated by an interaction with small momentum 
transfer. We also observe that for each plasma particle $e$, $\gamma$, $\tilde{e}$, 
$\tilde{\gamma}$ there is a process in which the cross section is independent of momentum 
transfer. 

\begin{table}
\caption{\label{tab-corr-binray} Cross sections of binary processes}
\begin{ruledtabular}
\begin{tabular}{m{.5cm} m{3.5cm} m{5cm} m{4cm}}
$n^0$ & \hspace{0.4cm} process & \hspace{0.9cm} diagrams 
& \hspace{0.7cm} cross section $\frac{d\sigma}{dt}$
\\ \\
         1& $e^{\mp}e^{\mp} \longrightarrow e^{\mp}e^{\mp}$ &
        \includegraphics[scale=.25]{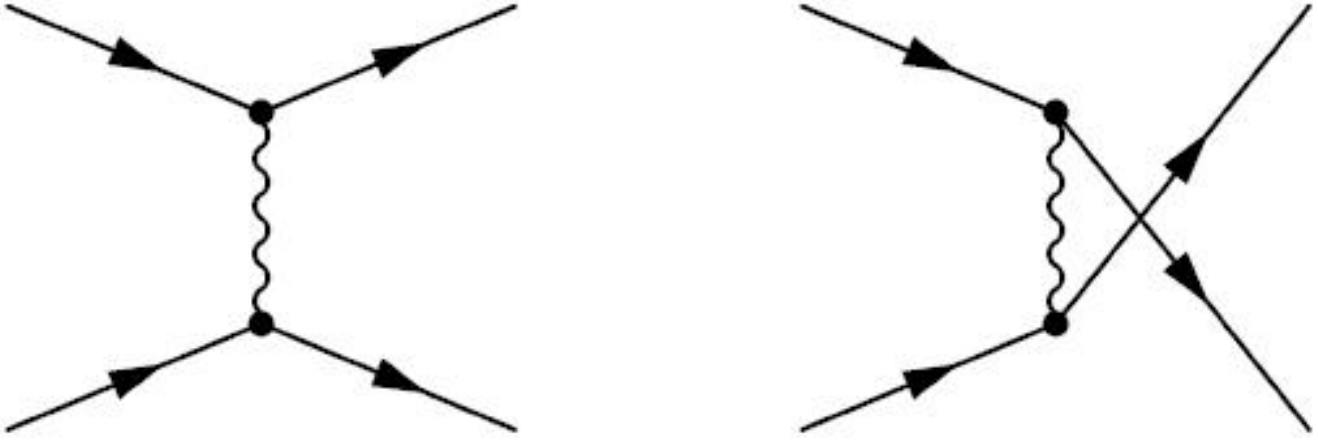} &
        $\frac{2\pi \alpha^2}{s^2} \big(\frac{s^2+u^2}{t^2}+\frac{s^2+t^2}{u^2}+\frac{2s^2}{tu}
        \big)$
        \\ \\
        2& $e^\pm e^\mp \longrightarrow  e^\pm e^\mp$ &
        \includegraphics[scale=.25]{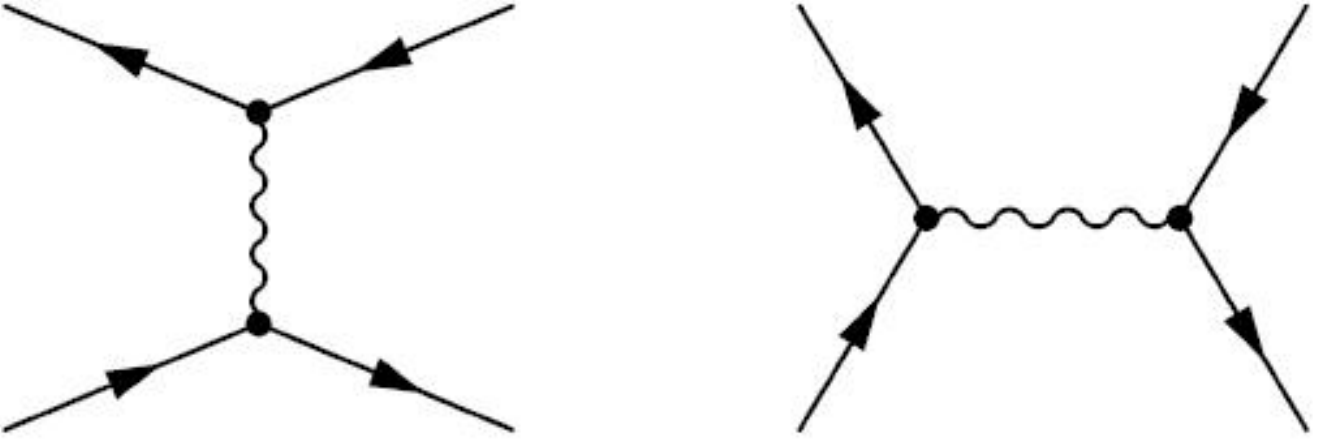} &
        $\frac{2\pi \alpha^2}{s^2} \big(\frac{s^2+u^2}{t^2}+\frac{u^2+t^2}{s^2}+\frac{2u^2}{ts}
        \big)$
        \\ \\
        3& $\gamma e^\mp \longrightarrow  \gamma e^\mp$ &
        \includegraphics[scale=.25]{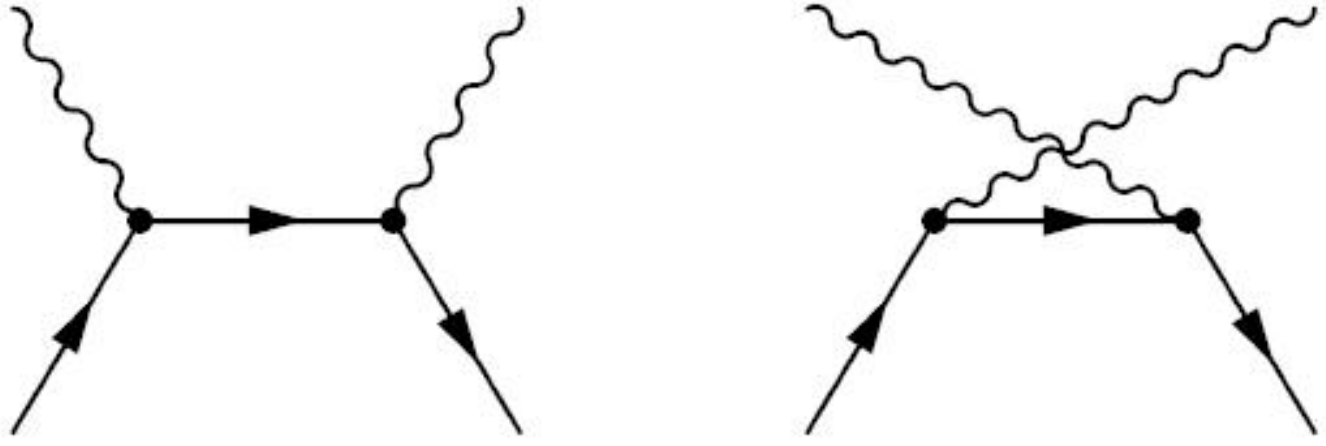}&
        $-\frac{2\pi \alpha^2}{s^2} \big(\frac{s}{u}+\frac{u}{s}
        \big)$
        \\ \\
        4& $e^\pm e^\mp \longrightarrow  \gamma \gamma $ &
        \includegraphics[scale=.25]{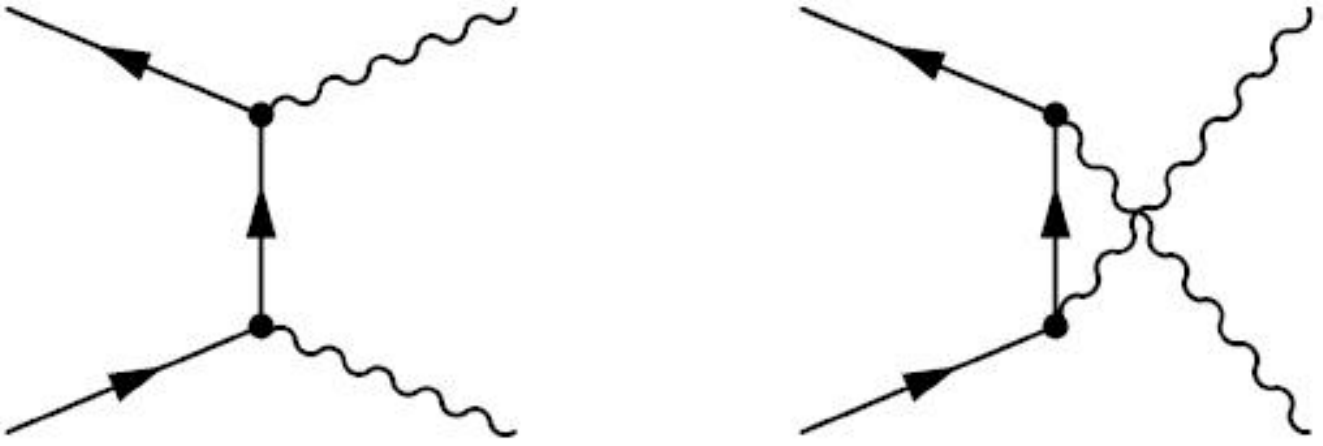}&
        $\frac{2\pi \alpha^2}{s^2} \big(\frac{t}{u}+\frac{u}{t}
        \big)$
        \\ \\
        5& $\gamma \gamma  \longrightarrow e^\mp e^\pm $ &
        \includegraphics[scale=.25]{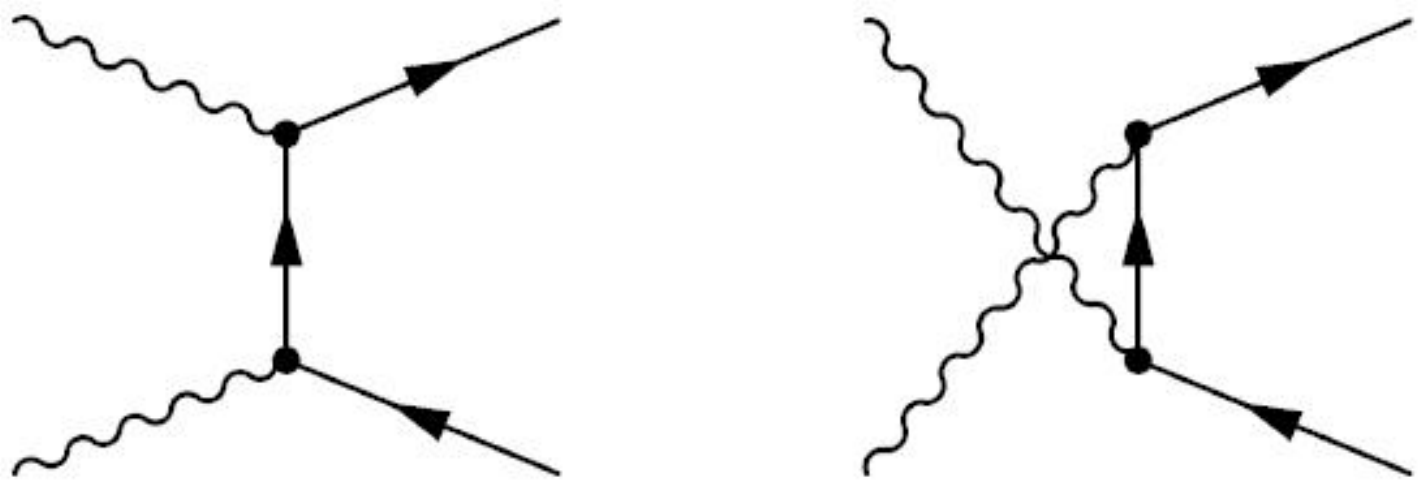}&
        $\frac{2\pi \alpha^2}{s^2} \big(\frac{t}{u}+\frac{u}{t}
        \big)$
        \\ \\
        6& $\tilde{\gamma} e^{\mp}  \longrightarrow \tilde{\gamma} e^{\mp}$ &
        \includegraphics[scale=.25]{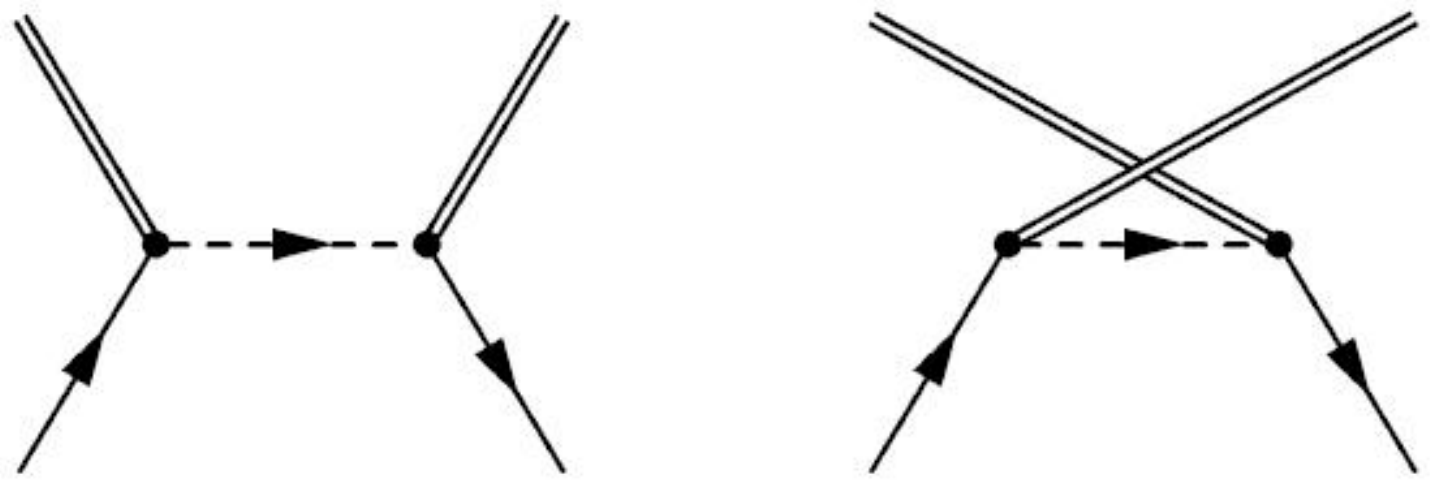}&
        $\frac{4\pi \alpha^2}{s^2}$
        \\ \\
        7& $e^{\pm} e^{\mp} \longrightarrow \tilde{\gamma}\tilde{\gamma}$ &
        \includegraphics[scale=.25]{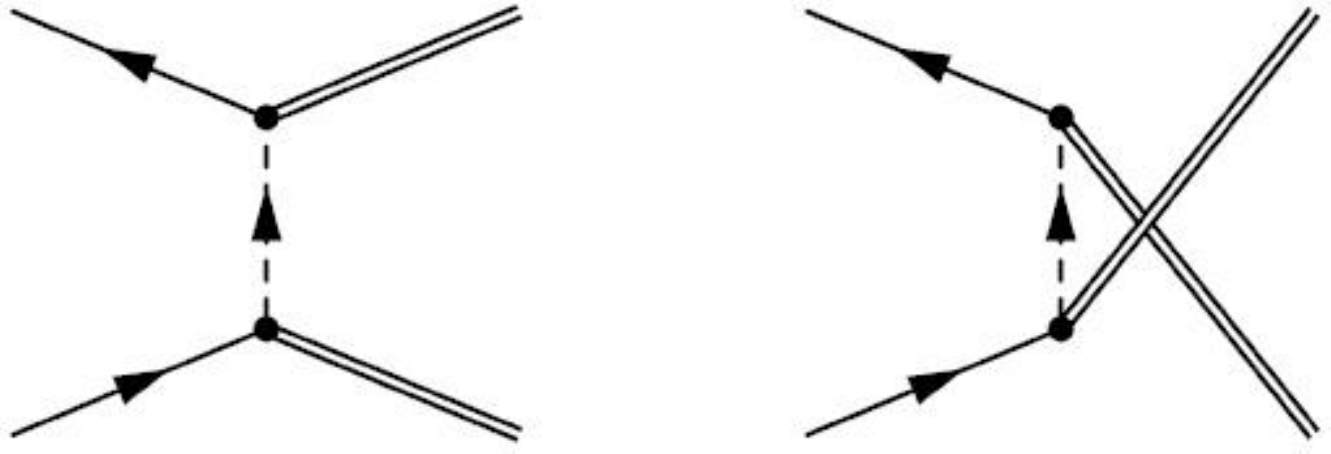}&
        $\frac{4\pi \alpha^2}{s^2}$
        \\ \\
        8& $\tilde{\gamma} \tilde{\gamma} \longrightarrow e^{\mp} e^{\pm}$ &
        \includegraphics[scale=.25]{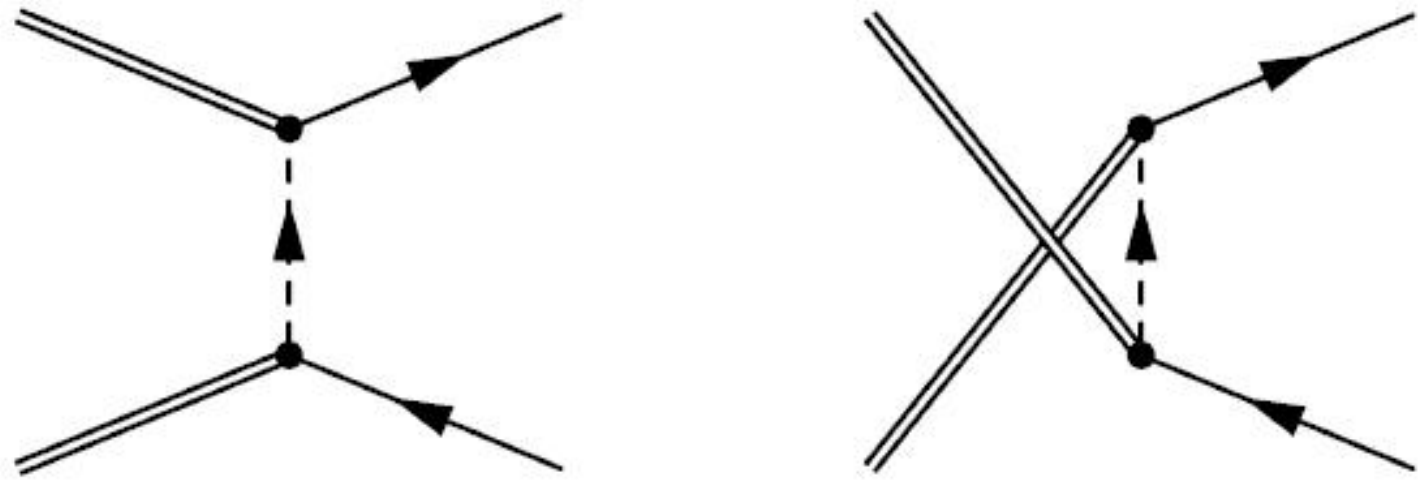}&
        $\frac{4\pi \alpha^2}{s^2}$
        \\ \\
        9& $\tilde{\gamma} e^{\mp} \longrightarrow \gamma \tilde{e}^{\mp}_{L,R} $ &
        \includegraphics[scale=.25]{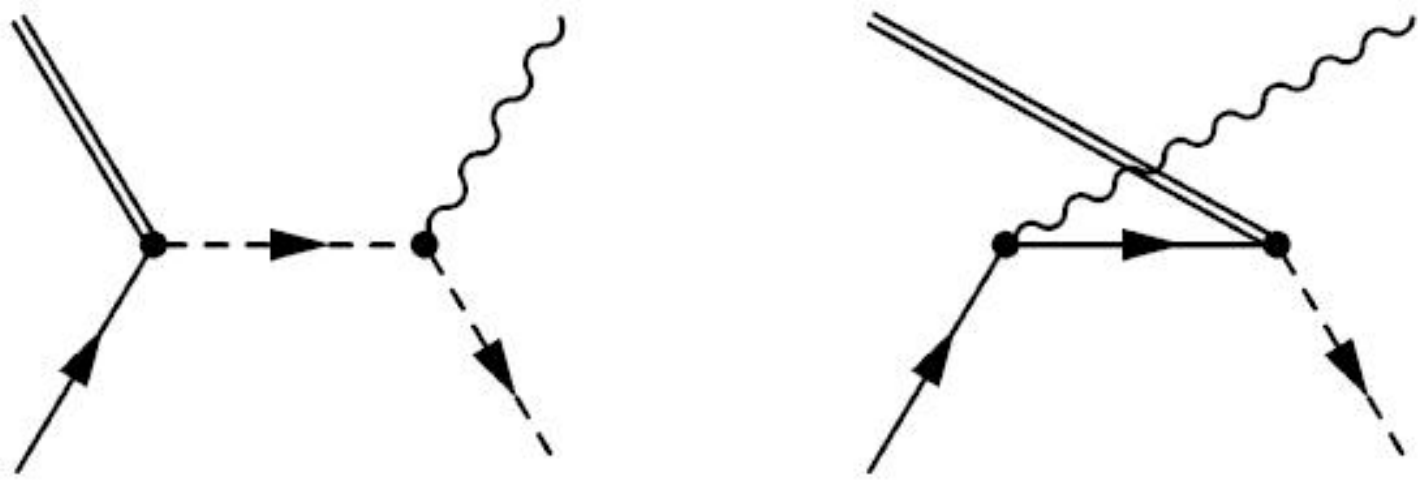}&
        $\frac{\pi \alpha^2}{s^2} \frac{t}{u}$
        \\ \\
        10& $ \gamma \tilde{e}^{\mp}_{L,R}  \longrightarrow  \tilde{\gamma} e^{\mp}$ &
        \includegraphics[scale=.25]{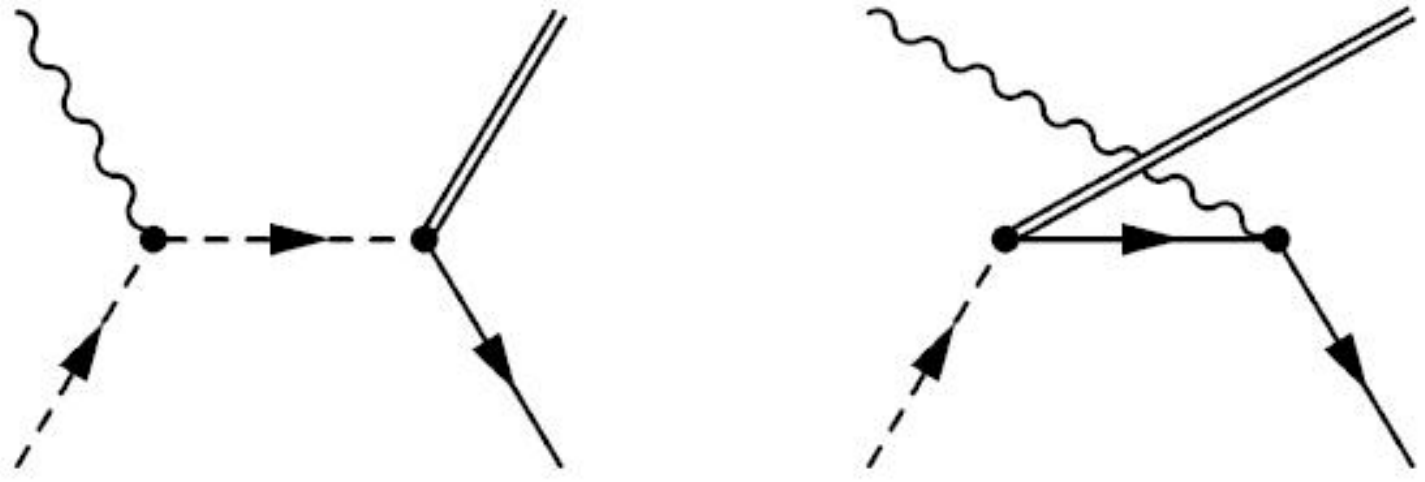}&
        $\frac{2\pi \alpha^2}{s^2} \frac{t}{u}$
         \\ \\
        11& $\gamma  e^{\mp} \longrightarrow \tilde{\gamma} \tilde{e}^{\mp}_{L,R} $ &
        \includegraphics[scale=.25]{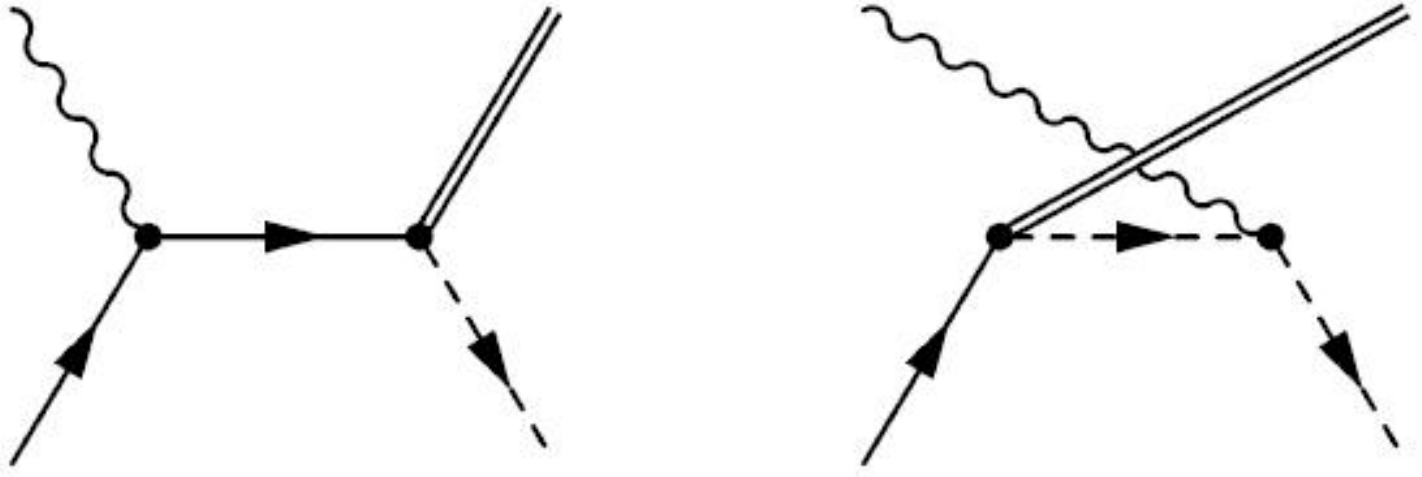}&
        $-\frac{2\pi \alpha^2}{s^2} \frac{t}{s}$
        \\ \\
        12& $\tilde{\gamma} \tilde{e}^{\mp}_{L,R}  \longrightarrow  \gamma e^{\mp} $ &
        \includegraphics[scale=.25]{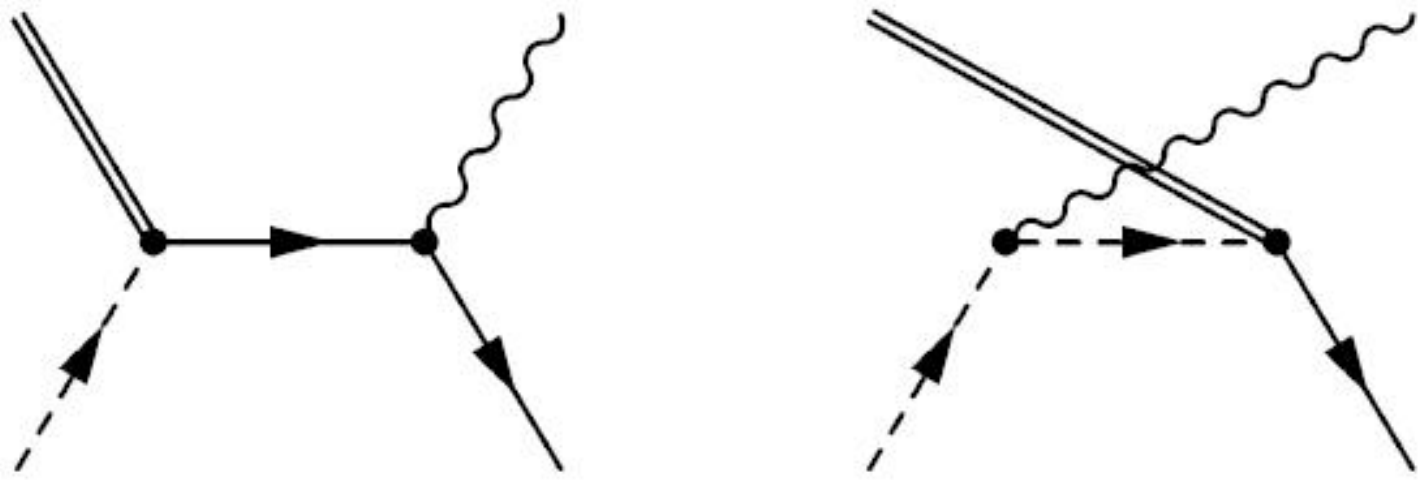}&
        $-\frac{2\pi \alpha^2}{s^2} \frac{t}{s}$
        \\ \\
        13& $\tilde{e}^{\pm}_{L,R} e^{\mp} \longrightarrow \tilde{\gamma} \gamma $ &
        \includegraphics[scale=.25]{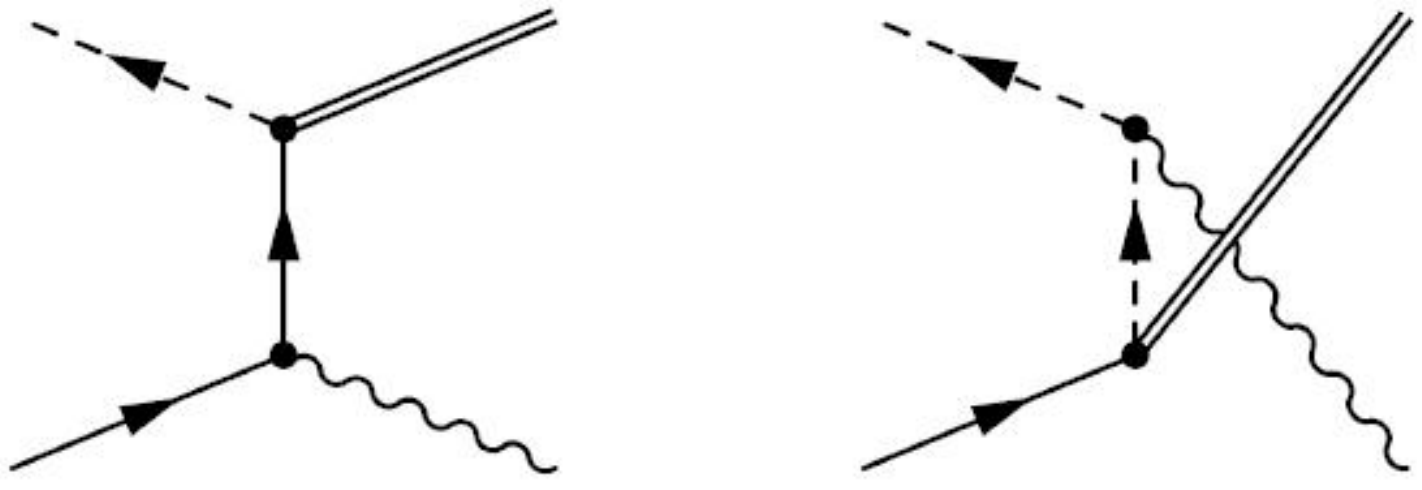}&
        $-\frac{2\pi \alpha^2}{s^2} \frac{s}{t}$
        \\ \\
        14& $\tilde{\gamma} \gamma \longrightarrow \tilde{e}^{\mp}_{L,R} e^{\pm} $ &
        \includegraphics[scale=.25]{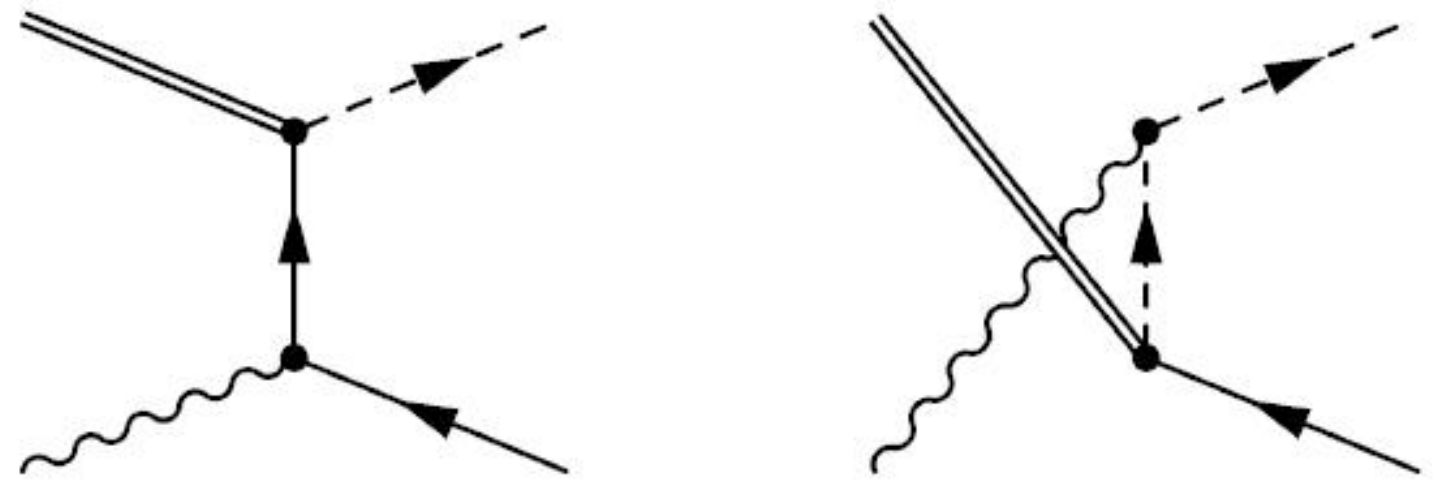}&
        $-\frac{\pi \alpha^2}{s^2} \frac{s}{t}$
         \\ \\
\end{tabular}
\end{ruledtabular}
\end{table}

\begin{table}
\begin{ruledtabular}
\begin{tabular}{m{.5cm} m{3.5cm} m{5cm} m{4cm}}
\\
        15& $\tilde{e}^{\mp}_{L,R}e^{\mp} \longrightarrow \tilde{e}^{\mp}_{L,R} e^{\mp}$ &
        \includegraphics[scale=.25]{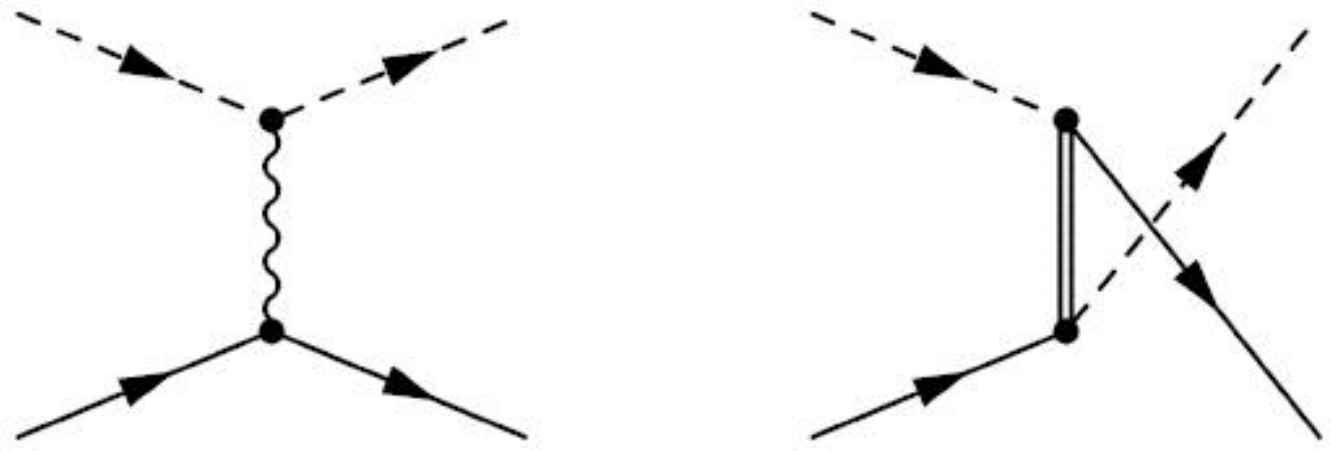}&
        $-\frac{2\pi \alpha^2}{s^2} \frac{s(s^2+u^2)}{ut^2}$
        \\ \\
        16& $\tilde{e}^{\pm}_{L,R} e^{\mp} \longrightarrow \tilde{e}^{\pm}_{L,R} e^{\mp}$ &
        \includegraphics[scale=.25]{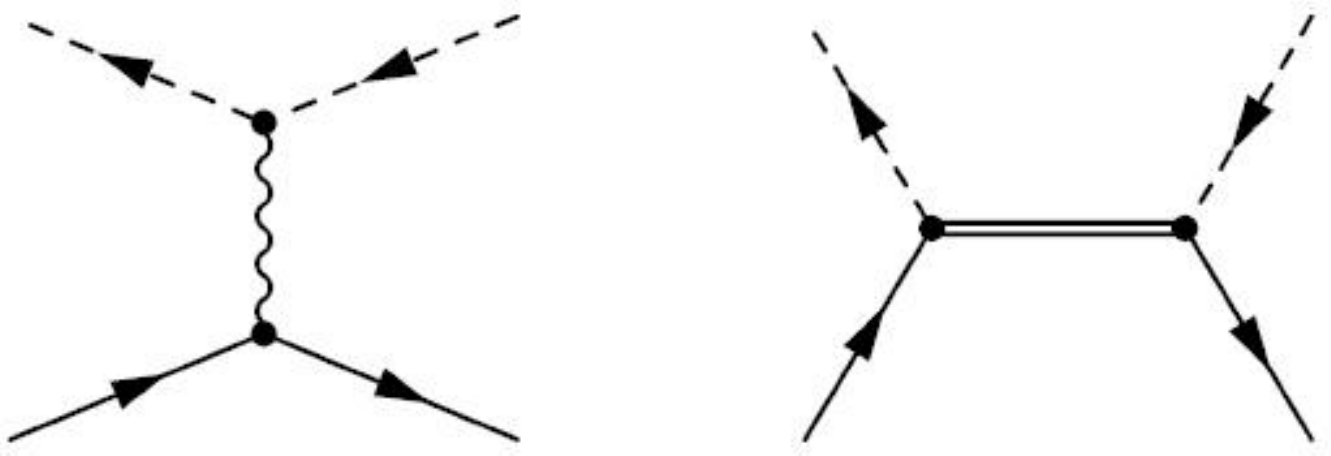}&
        $-\frac{2\pi \alpha^2}{s^2} \frac{u(s^2+u^2)}{st^2}$
        \\ \\
        17& $e^{\pm} e^{\mp} \longrightarrow \tilde{e}^{\pm}_{L,R} \tilde{e}^{\mp}_{L,R}$ &
        \includegraphics[scale=.25]{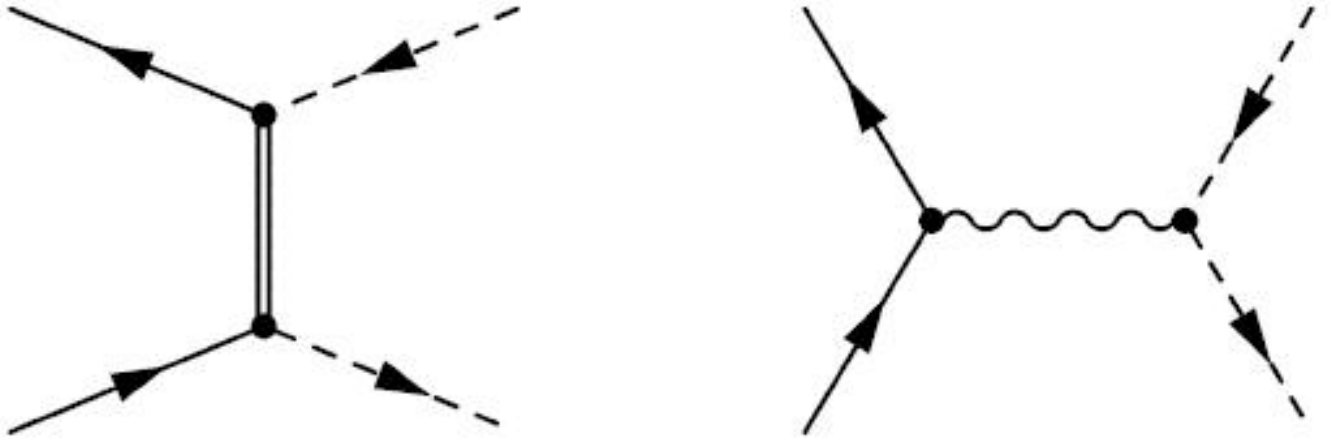}&
        $\frac{\pi \alpha^2}{s^2} \frac{u(t^2+u^2)}{ts^2}$
        \\ \\
        18& $\tilde{e}^{\pm}_{L,R} \tilde{e}^{\mp}_{L,R} \longrightarrow e^{\pm} e^{\mp}$ &
        \includegraphics[scale=.25]{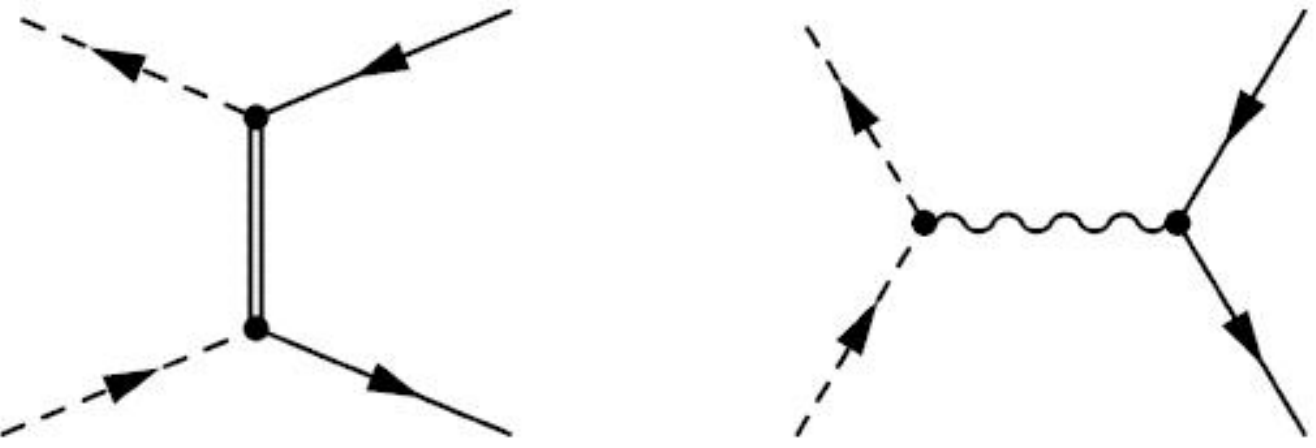}&
        $\frac{4\pi \alpha^2}{s^2} \frac{u(t^2+u^2)}{ts^2}$
        \\ \\
        19& $\tilde{e}^{\mp}_{L,R} \tilde{e}^{\mp}_{L,R} \longrightarrow \tilde{e}^{\mp}_{L,R} \tilde{e}^{\mp}_{L,R}$ &
        \includegraphics[scale=.25]{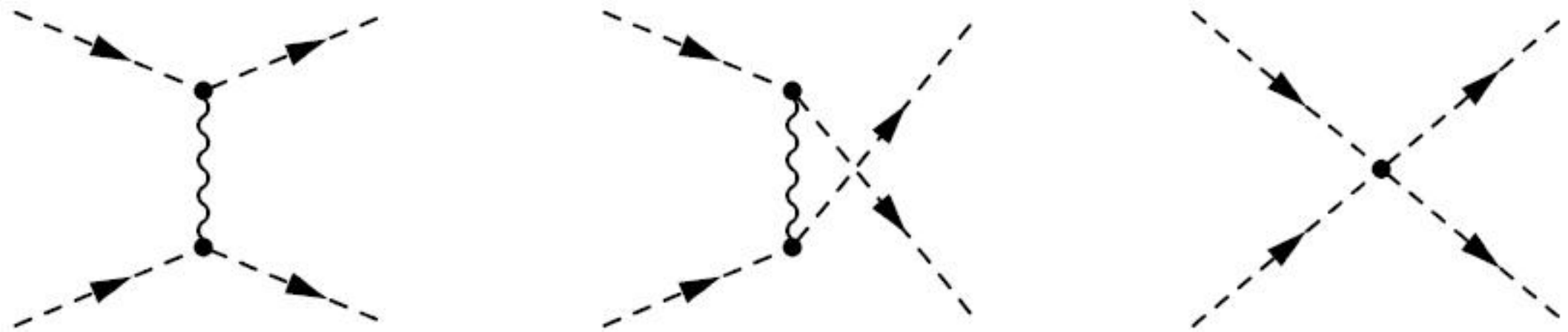}&
        $\frac{4\pi \alpha^2}{s^2}\big(\frac{u}{t} + \frac{t}{u}\big)^2$
        \\ \\
        20& $\tilde{e}^{\pm}_{L,R} \tilde{e}^{\mp}_{L,R} \longrightarrow \tilde{e}^{\pm}_{L,R} \tilde{e}^{\mp}_{L,R}$ &
        \includegraphics[scale=.25]{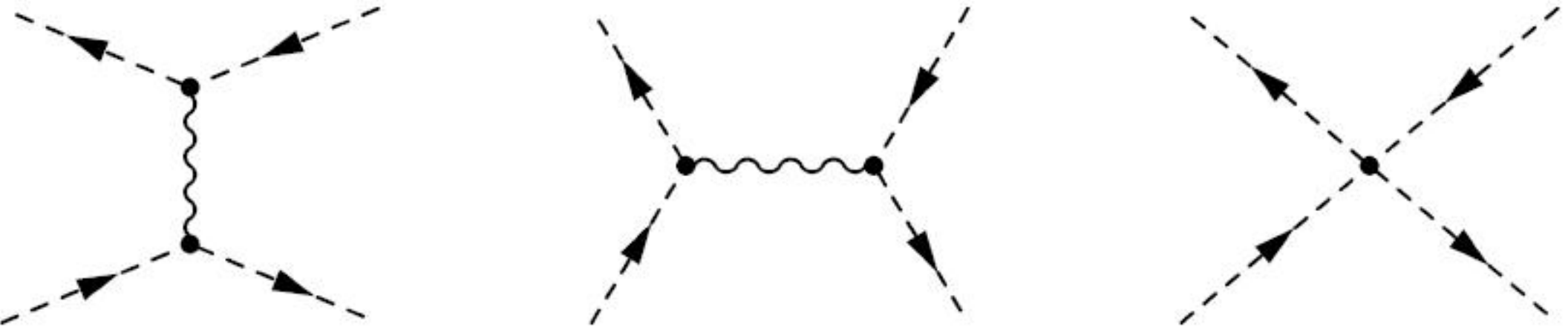}&
         $\frac{4\pi \alpha^2}{s^2}\big(\frac{s}{t}+\frac{t}{s} \big)^2$
        \\ \\
        21& $\tilde{e}^{\mp}_{L,R} \tilde{e}^{\mp}_{R,L} \longrightarrow \tilde{e}^{\mp}_{L,R} \tilde{e}^{\mp}_{R,L}$ &
        \includegraphics[scale=.25]{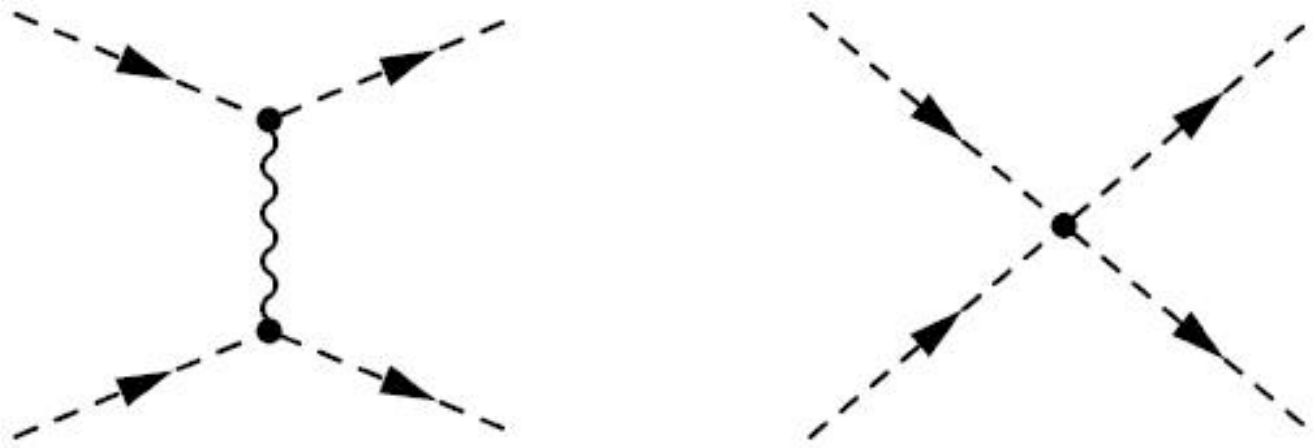}&
        $\frac{4\pi \alpha^2}{t^2}$
        \\ \\
        22& $\tilde{e}^{\pm}_{L,R} \tilde{e}^{\mp}_{R,L} \longrightarrow \tilde{e}^{\pm}_{L,R} \tilde{e}^{\mp}_{R,L}$ &
        \includegraphics[scale=.25]{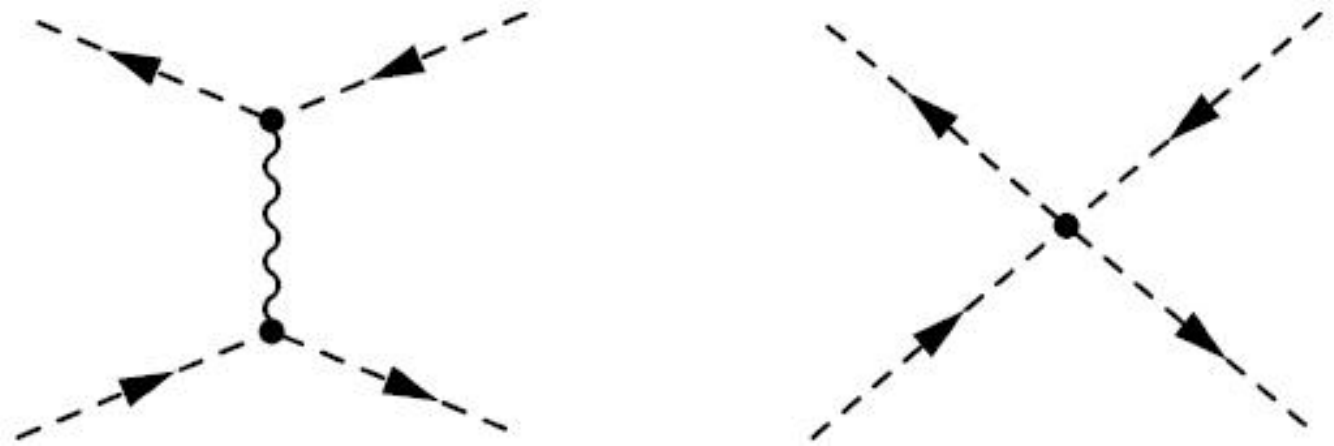}&
        $\frac{4\pi \alpha^2}{s^2} \frac{u^2}{t^2}$
        \\ \\
        23& $\tilde{e}^{\pm}_{L,R} \tilde{e}^{\mp}_{R,L} \longrightarrow \tilde{e}^{\pm}_{R,L} \tilde{e}^{\mp}_{L,R}$ &
        \includegraphics[scale=.25]{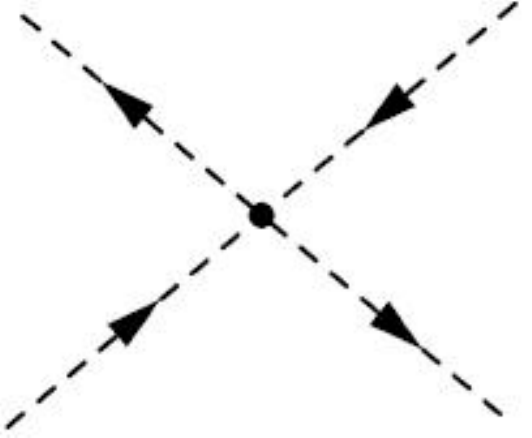} &
        $\frac{\pi \alpha^2}{s^2}$
        \\ \\
        24& $\tilde{e}^{\mp}_{L,R} \tilde{e}^{\mp}_{L,R}
        \longrightarrow \tilde{e}^{\mp}_{R,L} \tilde{e}^{\mp}_{R,L}$ &
        \includegraphics[scale=.25]{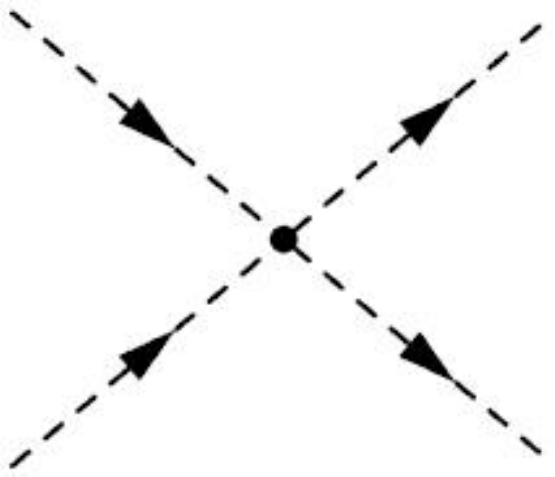}&
        $\frac{\pi \alpha^2}{s^2}$
        \\ \\
        25& $\gamma \tilde{e}^{\mp}_{L,R} \longrightarrow \gamma \tilde{e}^{\mp}_{L,R}$ &
        \includegraphics[scale=.25]{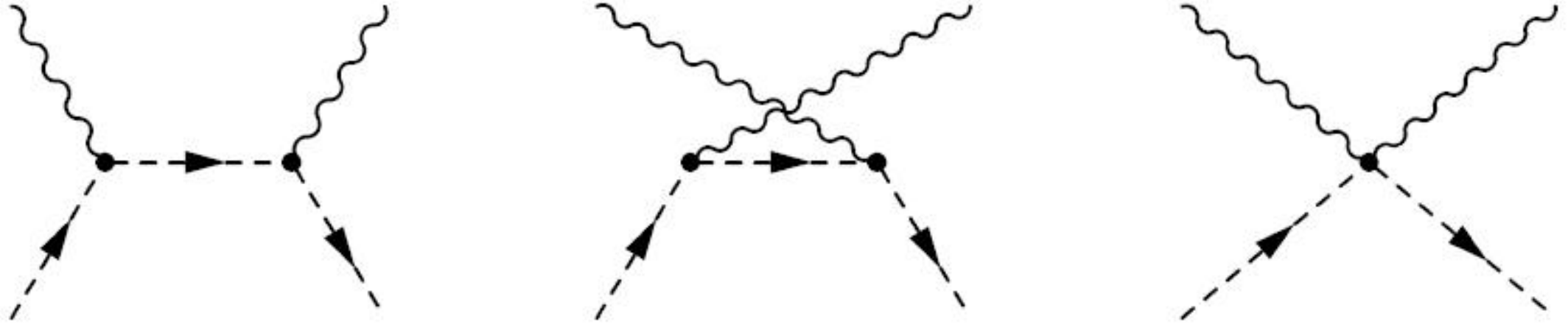}&
        $\frac{4\pi \alpha^2}{s^2}$
        \\ \\
        26& $\tilde{e}^{\pm}_{L,R} \tilde{e}^{\mp}_{L,R} \longrightarrow \gamma \gamma$ &
        \includegraphics[scale=.25]{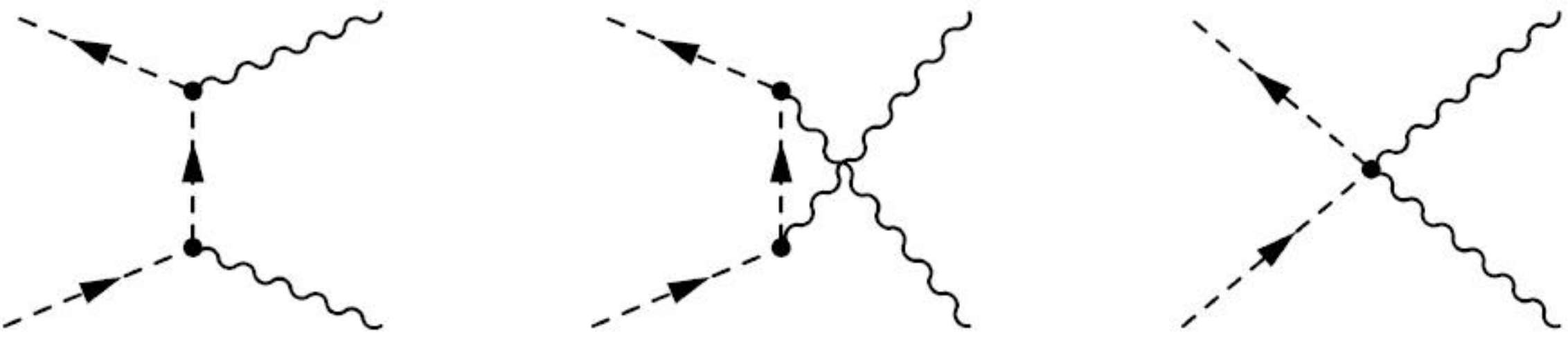}&
        $\frac{8\pi \alpha^2}{s^2}$
        \\ \\
        27& $\gamma \gamma \longrightarrow \tilde{e}^{\mp}_{L,R} \tilde{e}^{\pm}_{L,R}$ &
        \includegraphics[scale=.25]{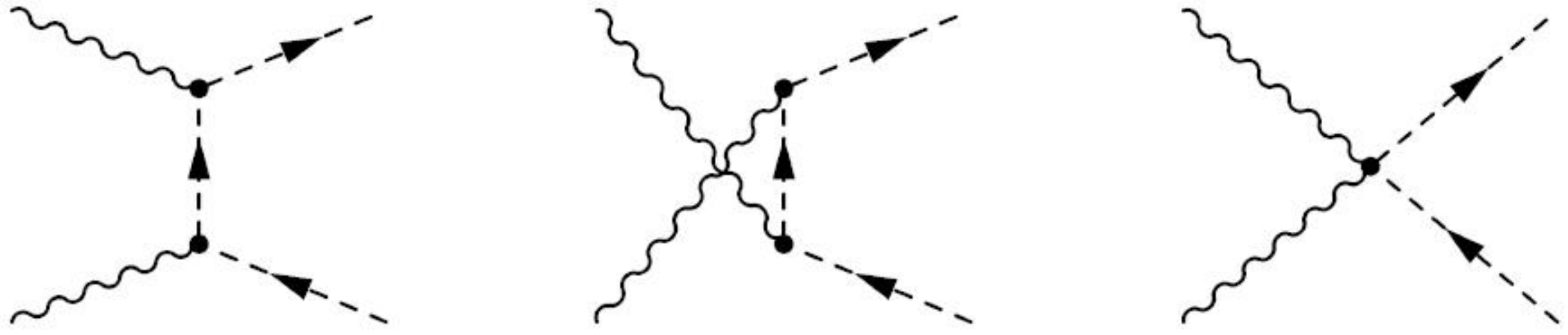}&
        $\frac{2\pi \alpha^2}{s^2}$
        \\ \\
        28& $\tilde{\gamma} \tilde{e}^{\mp}_{L,R} \longrightarrow \tilde{\gamma} \tilde{e}^{\mp}_{L,R} $ &
        \includegraphics[scale=.25]{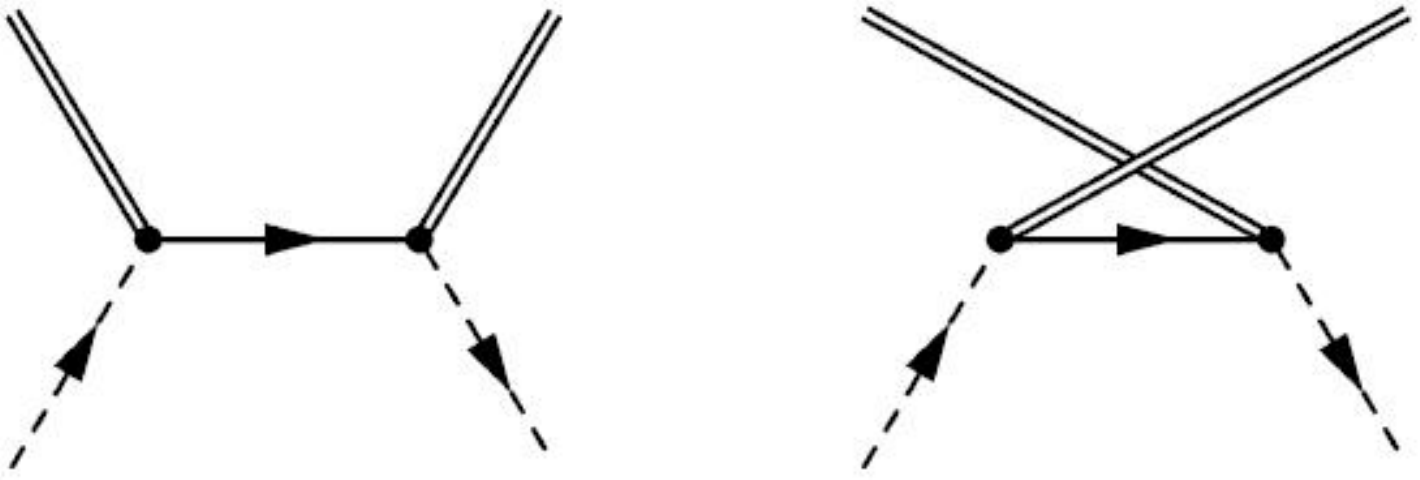}&
        $-\frac{2\pi \alpha^2}{s^2} \big(\frac{u}{s}+\frac{s}{u}
        \big)$
        \\ \\
\end{tabular}
\end{ruledtabular}
\end{table}

\begin{table}
\begin{ruledtabular}
\begin{tabular}{m{.5cm} m{3.5cm} m{5cm} m{4cm}}
\\
        29& $\tilde{e}^{\pm}_{L,R} \tilde{e}^{\mp}_{L,R} \longrightarrow \tilde{\gamma} \tilde{\gamma}$ &
        \includegraphics[scale=.25]{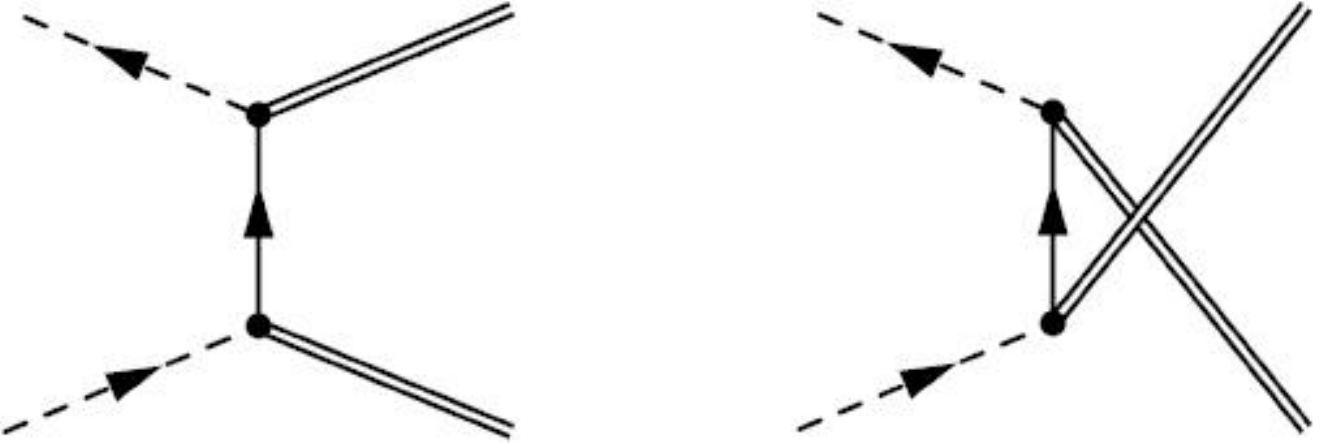}&
        $\frac{4\pi \alpha^2}{s^2} \big(\frac{u}{t}+\frac{t}{u}
        \big)$
        \\ \\
        30& $\tilde{\gamma} \tilde{\gamma} \longrightarrow \tilde{e}^{\pm}_{L,R} \tilde{e}^{\mp}_{L,R}$ &
        \includegraphics[scale=.25]{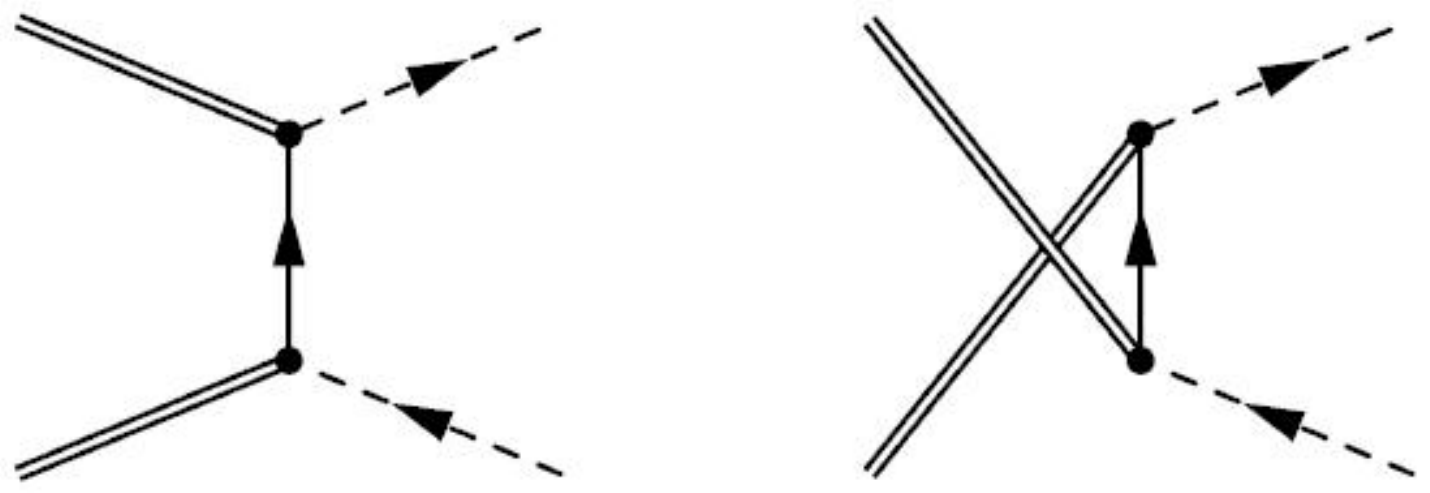}&
        $\frac{\pi \alpha^2}{s^2} \big(\frac{u}{t}+\frac{t}{u}
        \big)$
        \\ \\
        31& $\tilde{e}^{\pm}_{L,R} e^{\mp} \longrightarrow e^{\pm} \tilde{e}^{\mp}_{R,L}$ &
        \includegraphics[scale=.25]{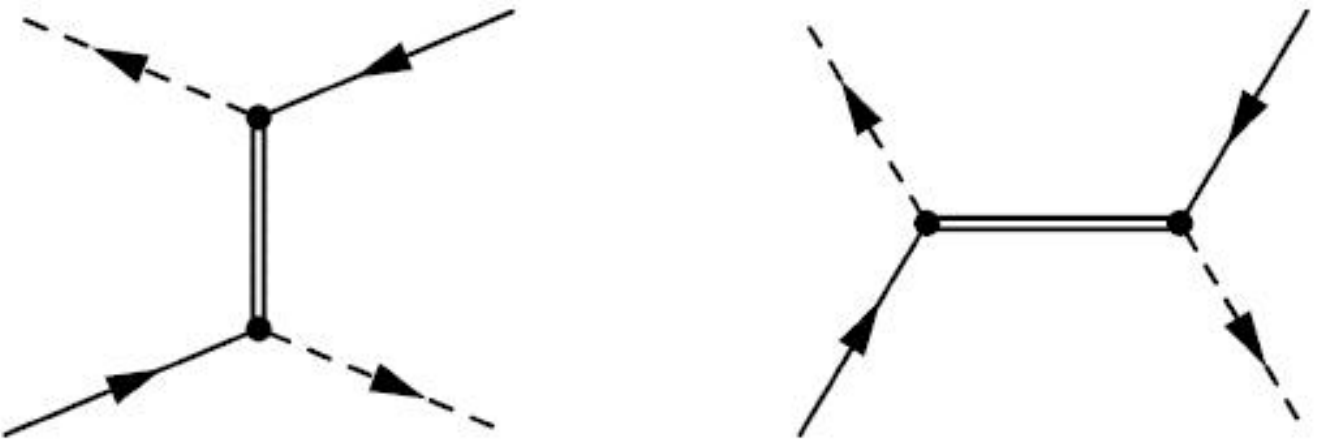}&
        $-\frac{2\pi \alpha^2}{s^2}\big( \frac{s}{t} + \frac{t}{s} \big)$
        \\ \\
        32& $e^{\mp} e^{\mp} \longrightarrow \tilde{e}^{\mp}_{L,R} \tilde{e}^{\mp}_{R,L} $ &
        \includegraphics[scale=.25]{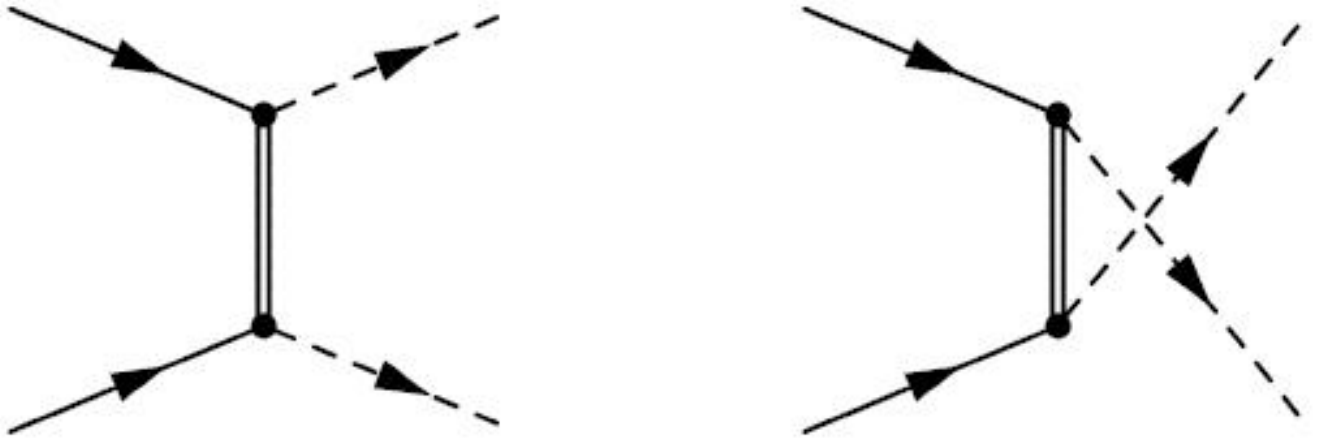} &
        $\frac{\pi \alpha^2}{s^2}\big( \frac{u}{t} + \frac{t}{u} \big)$
        \\ \\
        33& $\tilde{e}^{\mp}_{L,R} \tilde{e}^{\mp}_{R,L} \longrightarrow e^{\mp} e^{\mp}$ &
        \includegraphics[scale=.25]{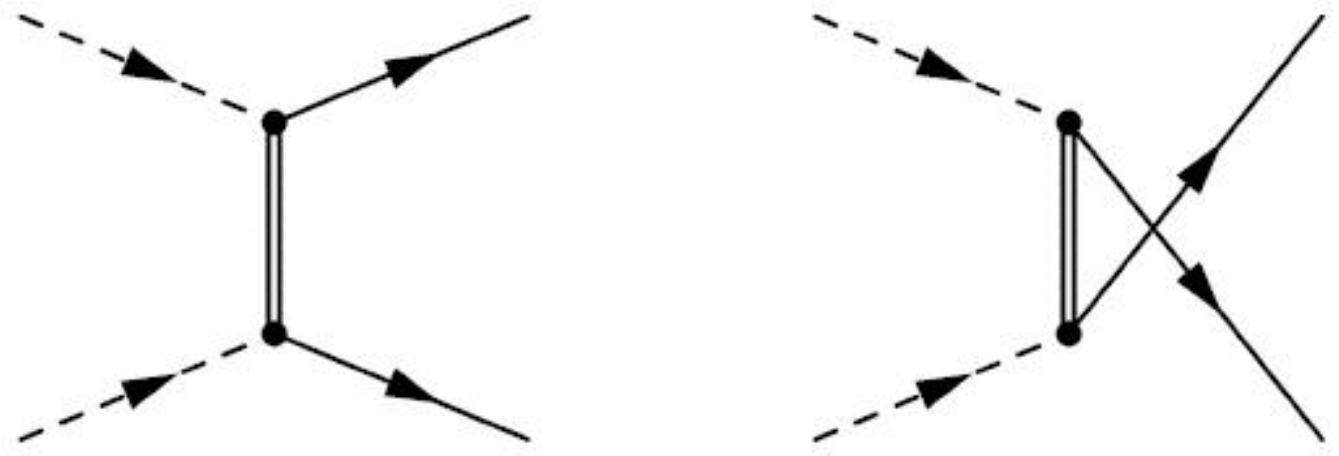}&
        $\frac{4\pi \alpha^2}{s^2}\big( \frac{u}{t} + \frac{t}{u} \big)$
        \\ \\
\end{tabular}
\end{ruledtabular}
\end{table}

\section{Energy loss}
\label{sec-e-loss}

As already mentioned, for each type of plasma particle there is a binary process with 
a cross section independent of $t$ or $u$. The matrix element of such a process is 
simply a number. We compute here the energy loss caused by such an interaction. 
To be specific, we consider a high energy selectron traversing an equilibrium ${\cal N} =1$ 
SUSY QED plasma. The selectron interacts with plasma particles of all types but we take 
into account only scattering on photons. This is only for demonstration but in general
all processes, which contribute additively to the energy loss, must be included.

The selectron initial four-momentum is denoted as 
$p^\mu = (E, {\bf p})$ with $E \equiv |{\bf p}|$ and that of a plasma photon as
$p^\mu_1 = (E_1, {\bf p}_1)$. The final four-momenta of the selectron and photon are, 
respectively, $p'^\mu = (E', {\bf p}')$ and  ${p'}^\mu_1 = (E'_1, {\bf p}'_1)$. 
The energy loss of the selectron per unit length is then
\be
\label{e-loss-def}
\frac{dE}{dx}= -\int d\Gamma (E -  E'),
\ee
where $d\Gamma$ is the interaction rate given as
\be
\label{rate-def}
d\Gamma = |\mathcal{M}|^2 \frac{n(\mathbf{p}_1) }{16 E E' E_1 E'_1} 
\, \frac{d^3p_1}{(2\pi)^3} \, \frac{d^3p'}{(2\pi)^3} \,
 \frac{d^3p'_1}{(2\pi)^3}\, (2\pi)^4 \,   \delta^{(4)}(p + p_1 -p' - p'_1) ;
\ee
$n(\mathbf{p}_1)$ is the distribution function of plasma photons and $\mathcal{M}$ 
denotes the scattering amplitude. We have neglected here the quantum factor 
$n(\mathbf{p}_1') +1$ which is important when the momentum of final state 
photon is of order of plasma temperature. Since the scattering process under 
consideration leads to a sizable momentum transfer and we are mostly interested 
in energy loss of a highly energetic particle, the factor can be safely  ignored. 

When $\mathcal{M}$ describes a scattering driven by a one-photon exchange,
the formula (\ref{e-loss-def}) with the rate (\ref{rate-def}) leads to an infinite
result due to the long range nature of electromagnetic interaction. The problem 
is cured by including the effect of screening in a plasma medium. In the case of 
photon-selectron scattering the matrix element equals $|\mathcal{M}|^2 = 4 e^4$ 
and it does not need any modification to provide a finite energy loss. 

Substituting the interaction rate (\ref{rate-def}) with $|\mathcal{M}|^2 = 4 e^4$ into 
Eq.~(\ref{e-loss-def}) and performing the trivial integration over ${\bf p}'_1$, we obtain 
\be
\label{e-loss1}
\frac{dE}{dx}= -\frac{e^4}{4}\int \frac{d^3p_1}{(2\pi)^3}
\frac{d^3q}{(2\pi)^3} \frac{n(\mathbf{p}_1)(E-E')}{E^2 E_1 E'_1 q} 
\: 2\pi \delta(\cos\theta - \overline{\cos \theta} ) ,
\ee
where ${\bf q} \equiv {\bf p}' - {\bf p}={\bf p}_1 - {\bf p}'_1$ is the momentum transfer 
and $q\equiv |{\bf q}|$; $\theta$ is the angle between the vectors ${\bf p}$ and 
${\bf q}$ and $\overline{\cos \theta}$ is the solution of the energy conservation equation
\be
\label{cos-bar}
\overline{\cos \theta} = \frac{(E+E_1 - E'_1)^2 - E^2 - q^2}{2 q E}
\ee
provided $-1 \le \overline{\cos \theta} \le 1$.

Now we make use of the assumption that the plasma is in thermal equilibrium and therefore 
is isotropic. As a result, the momentum distribution of plasma photons depends on ${\bf p}_1$  
only through $E_1$ and we write it as $n(E_1)$. Consequently, the energy loss is independent 
of the orientation of the momentum $\mathbf{p}$. Therefore, following \cite{Braaten:1991jj} 
we average the formula (\ref{e-loss1}) over the orientation of $\mathbf{p}$ with respect 
to ${\bf q}$ and we get 
\be
\label{e-loss2}
\frac{dE}{dx}= \int \frac{d\Omega}{4\pi} \frac{dE}{dx}
= -\frac{e^4}{2^8 \pi^5}\int d^3p_1 d^3q 
\, \frac{n(E_1) (E-E')}{E^2 E_1 E'_1 q}.
\ee
We write down the integral over ${\bf q}$ in spherical coordinates where the axis $z$ 
is along the momentum ${\bf p}_1$. Then, the integral over orientation of ${\bf p}_1$
is trivial and one obtains
\be
\label{e-loss-3}
\frac{dE}{dx}= -\frac{e^4}{2^5 \pi^3}
\int_0^\infty dE_1 E_1^2
\int_{q_{\rm min}}^{q_{\rm max}} dq q^2 
\int_{(\cos \theta_1)_{\rm min}}^{(\cos \theta_1)_{\rm max}}
d(\cos\theta_1)
\frac{n(E_1)\, (E - E')}{E^2 E_1 E'_1 q},
\ee
where $\theta_1$ is the angle between the vectors ${\bf p}_1$ and ${\bf q}$. The integration 
limits must be chosen in such a way that the energy conservation is satisfied. 
Instead of $\cos \theta_1$ it appears more convenient to use the variable 
$\omega \equiv E -E' = \sqrt{E_1^2 -2 E_1 q \cos \theta_1 +q^2} - E_1$.
Then, the expression (\ref{e-loss-3}) can be written in the form
\be
\label{e-loss-4}
\frac{dE}{dx} = -\frac{e^4}{2^5 \pi^3 E^2}\int_0^\infty d E_1n(E_1) 
\int_{q_{\rm min}}^{q_{\rm max}}dq 
\int_{\omega_{\rm min}}^{\omega_{\rm max}}
d\omega \, \omega .
\ee

To find the integration limits we express $\overline{\cos \theta}$, which is given
by Eq.~(\ref{cos-bar}), through the variable $\omega$ and we demand that
$-1 \le \overline{\cos \theta} \le 1$ keeping in mind that $-E_1 \le \omega \le E$. 
Then, a somewhat lengthy but elementary analysis leads to the expression
\ba
 \frac{dE}{dx} &=& -\frac{e^4}{2^5 \pi^3 \, E^2}
\Big\{ \int_0^{E} {\rm d}E_1 n(E_1) \Big[
\int_0^{E_1} d q \int_{-q}^{q} d \omega \, \omega
+  \int_{E_1}^E d q \int_{q -2E_1}^{q}  d\omega \, \omega 
+  \int_E^{E+E_1} dq \int_{q -2E_1}^{2E-q} d \omega \, \omega \Big]
\nn \\
&&\;\;\;\;\;\;\;\;\;\;\;\;\;\;\;
+ \int_E^{\infty} d E_1 n(E_1) 
\Big[ \int_0^E d q \int_{-q}^{q} d \omega \, \omega 
+  \int_E^{E_1} dq \int_{-q}^{2E - q} d\omega \, \omega
+ \int_{E_1}^{E+E_1} d q \int_{q -2E_1}^{2E - q} d \omega \, \omega 
\Big] \Big\},
\label{e-loss-5}
\ea
which after performing simple integrations over $q$ and $\omega$ gives
\be
\label{e-loss-n(E)}
\frac{d E}{dx} = -\frac{e^4}{2^5 \pi^3 E}
\int_0^{\infty}  d E_1 n(E_1) \big( E E_1 - E_1^2\big).
\ee

To check correctness of rather complicated integration domain in Eq.~(\ref{e-loss-n(E)}), one 
observes that the integral (\ref{e-loss-5}) becomes simple when $\omega \equiv E -E'$ is 
replaced by unity.  Then, the integral 
\be
\int \frac{d^3p'}{(2\pi)^3 2E'} \,
 \frac{d^3p'_1}{(2\pi)^3 2E_1'}\, (2\pi)^4 \,   \delta^{(4)}(p + p_1 -p' - p'_1)  = \frac{1}{8\pi}
\ee 
is Lorentz invariant and it can be easily computed in the center-of-mass frame. We have
reproduced this result in an arbitrary frame performing the integration over the domain in
$q-\omega$ space shown in Eq.~(\ref{e-loss-5}).

The energy distribution of photons in equilibrium plasma is of Bose-Einstein form
\be
\label{B-E}
n(E) = \frac{2}{e^{\frac{E}{T}}-1},
\ee
where the factor of 2 takes into account two photon polarizations and $T$ is the plasma 
temperature.  Substituting the distribution (\ref{B-E}) into Eq.~(\ref{e-loss-n(E)}),
one finds 
\be
\label{e-loss-final}
\frac{dE}{d x} = -\frac{e^4}{2^5 3 \pi}\, T^2 
\Big[1 - \frac{12 \zeta(3)}{\pi^2}  \frac{T}{E} \Big] ,
\ee
where $\zeta(z)$ is the zeta Riemann function and $\zeta(3) \approx 1.202$.

In the limit $E \gg T$ we have the result
\be
\label{e-loss-BigE}
\frac{d E}{d x}=   -\frac{e^4}{2^5 3 \pi}\, T^2 , 
\ee
which should be confronted with the energy loss of an energetic muon in ultrarelativistic 
electromagnetic plasma of electrons, positrons and photons \cite{Braaten:1991jj}
\be
\label{e-loss-QED}
\frac{dE}{dx}= -\frac{e^4}{48\pi^3}\, T^2 \Big( \ln\frac{E}{eT}+2.031\Big).
\ee
As seen, the formulas (\ref{e-loss-BigE}, \ref{e-loss-QED}) are similar to each other
up the logarithm term which is discussed later on. The similarity is rather surprising 
if one realizes how different are the differential cross sections of interest.  Let us discuss 
why it happens.

The energy loss can be estimated as \cite{Peigne:2008wu} 
$\frac{dE}{d x} \sim \langle \Delta E \rangle / \lambda$, where $\langle \Delta E \rangle$ 
is the typical change of particle's energy in a single collision and $\lambda$ is the particle's 
mean free path given as $\lambda^{-1} = \rho \, \sigma$ with $\rho \sim T^3$ being the 
density of scatterers and $\sigma$ denoting the cross section. For the differential cross
section $\frac{ d\sigma}{dt} \sim e^4/s^2$, the total cross section is $\sigma \sim e^4/s$. When
a highly energetic particle with energy $E$ scatters on massless plasma particle, $s \sim ET$ and 
consequently  $\sigma \sim e^4/(ET)$. The inverse mean free path is thus estimated as 
$\lambda^{-1} \sim e^4 T^2/E$.  When the scattering process is independent of momentum transfer,
$\langle \Delta E \rangle$ is of order $E$ and we finally find $-\frac{dE}{d x} \sim e^4 T^2$. 
When compared to the case of Coulomb scattering, the energy transfer in a single collision is 
much bigger but the cross section is smaller in the same proportion. Consequently, the two 
interactions corresponding to very different differential cross sections lead to very similar energy 
losses.  The authors of \cite{Dusling:2009df} arrived to the analogous conclusion discussing 
viscous corrections to the distribution function caused by the collisions driven by a one-gluon 
exchange or by a $\phi^4$ contact interaction.

The Coulomb energy-loss formula (\ref{e-loss-QED})  differs from (\ref{e-loss-BigE}) by the 
logarithm term which comes from the integration over the momentum transfer from the minimal 
($q_{\rm min}$) to maximal ($q_{\rm max}$) value. The latter one is of order of energy of the 
test particle ($q_{\rm max} \sim E$).  In vacuum $q_{\rm min}=0$ and consequently  the integral, 
which equals $\ln (q_{\rm max}/q_{\rm min})$, diverges. In a plasma medium the long range 
Coulomb forces are screened and $q_{\rm min}$ is of order of Debye mass which in 
ultrarelativistic plasma is roughly $eT$. Thus, the logarithm term gets the form as in 
Eq.~(\ref{e-loss-QED}).

\section{Broadening of transverse momentum}
\label{sec-qhat}

We consider here a second transport characteristic of an equilibrium ${\cal N} =1$ SUSY 
QED plasma which is the momentum broadening of an energetic selectron due to its interaction 
with plasma photons. The quantity, which is usually denoted as $\hat q$, 
determines the magnitude of radiative energy loss of a highly energetic particle in a plasma medium
\cite{Baier:1996sk}. It is defined  as
\be
\label{qhat-def}
\hat q = \int d\Gamma \, q_T^2,
\ee
where $d\Gamma$ is, as previously, the interaction rate and $q_T$ is the momentum transfer 
to the selectron which is perpendicular to the selectron initial momentum. 

Since $\hat q$ is computed in exactly the same way as the energy loss, it can be
obtained by replacing $E' - E$ by $q_T^2$ in the formulas from the previous section. 
Then, instead of equation (\ref{e-loss-n(E)}) one finds
\be
\label{qhat-n(E)}
\hat{q} = \frac{e^4}{2^4 3 \pi^3 E}
\int_0^{\infty}  d E_1 n(E_1) \Big[E E_1^2 + \frac{2}{3}E_1^3 \Big].
\ee

With the momentum distribution of plasma photons of the Bose-Einstein form (\ref{B-E}),
Eq.~(\ref{qhat-n(E)}) gives
\be
\label{qhat-final}
\hat{q} = \frac{e^4}{12 \pi^3}\, T^3 
\bigg[ \zeta(3)  + \frac{\pi^4}{45}  \frac{T}{E}\bigg] .
\ee

When the momentum broadening is caused by scattering driven by one-photon 
exchange, $\hat{q}$ is of the order $e^4 \ln(1/e) \, T^3$ \cite{Arnold:2008vd}. 
Therefore, we conclude that the momentum broadening and consequently the radiative 
energy loss of a highly energetic particle in the SUSY QED plasma is similar (up the 
logarithm term) to that in the electromagnetic plasma of electrons, positrons and photons.
The logarithm occurs for the same reason as in the case of energy-loss formula.

\section{Summary and conclusions}
\label{sec-conclude}

After studying collective excitations of ${\cal N} =1$ SUSY QED  plasma and finding them
very similar to that of electromagnetic plasma of electrons, positrons and photons 
\cite{Czajka:2010zh}, we have focused our attention on collisional processes. First of all 
we have computed cross sections of all binary interactions which occur in the ${\cal N} =1$ 
SUSY QED  plasma at the lowest nontrivial order of $\alpha$. We have found a class of 
processes, Compton scattering on selectrons for example, the cross sections of which are 
independent of momentum transfer. These processes, in contrast to those characteristic for 
QED plasma, are not dominated by interactions with small momentum transfer and there exists
such a process for a plasma particles of every type. One can suspect that due to these processes, 
transport properties of a supersymmetric system are different than those of its non-supersymmetric  
counterpart.  We have derived the formulas for energy loss and momentum broadening of an energetic 
particle, showing that the ${\cal N} =1$ SUSY QED  plasma is actually surprisingly similar to the 
QED plasma. 

\section*{Acknowledgments}

We are very grateful to Mike Strickland for critical reading of the manuscript
and various comments. This work was partially supported by the ESF Human Capital 
Operational Program under grant 6/1/8.2.1/POKL/2009 and by the Polish Ministry 
of Science and Higher  Education under grants N~N202~204638 and 667/N-CERN/2010/0.



\begin{thebibliography}{99}

\bibitem{Czajka:2010zh}
A.~Czajka and St.~Mr\'owczy\'nski,
Phys.\ Rev.\  {\bf D83}, 045001 (2011).

\bibitem{Aharony:1999ti}
O.~Aharony, S.~S.~Gubser, J.~M.~Maldacena, H.~Ooguri, and Y.~Oz,
Phys.\ Rept.\  {\bf 323}, 183 (2000).

\bibitem{Son:2007vk}
D.~T.~Son and A.~O.~Starinets,
Ann.\ Rev.\ Nucl.\ Part.\ Sci.\  {\bf 57}, 95 (2007).

\bibitem{Janik:2010we}
R.~A.~Janik,
Lect.\ Notes Phys.\  {\bf 828}, 147 (2011).

\bibitem{CaronHuot:2006te}
S.~Caron-Huot, P.~Kovtun, G.~D.~Moore, A.~Starinets, and L.~G.~Yaffe,
JHEP {\bf 0612}, 015 (2006).

\bibitem{Huot:2006ys}
S.~C.~Huot, S.~Jeon, and G.~D.~Moore,
Phys.\ Rev.\ Lett.\  {\bf 98}, 172303 (2007).

\bibitem{CaronHuot:2008uh}
S.~Caron-Huot and G.~D.~Moore,
JHEP {\bf 0802}, 081 (2008).

\bibitem{Blaizot:2006tk}
J.~P.~Blaizot, E.~Iancu, U.~Kraemmer, and A.~Rebhan,
JHEP {\bf 0706}, 035 (2007).

\bibitem{Chesler:2006gr}
P.~M.~Chesler and A.~Vuorinen,
JHEP {\bf 0611}, 037 (2006).

\bibitem{Chesler:2009yg}
P.~M.~Chesler, A.~Gynther, and A.~Vuorinen,
JHEP {\bf 0909}, 003 (2009).

\bibitem{Baier:1996sk}
R.~Baier, Y.~L.~Dokshitzer, A.~H.~Mueller, S.~Peigne, and D.~Schiff,
Nucl.\ Phys.\  B {\bf 484}, 265 (1997).

\bibitem{Binoth:2002xg}
T.~Binoth, E.~W.~N.~Glover, P.~Marquard, and J.~J.~van der Bij,
JHEP {\bf 0205}, 060 (2002).

\bibitem{Braaten:1991jj}
E.~Braaten and M.~H.~Thoma,
Phys.\ Rev.\  {\bf D44}, 1298 (1991).

\bibitem{Peigne:2008wu}
S.~Peigne and A.~V.~Smilga,
Phys.\ Usp.\  {\bf 52}, 659 (2009).

\bibitem{Dusling:2009df}
K.~Dusling, G.~D.~Moore, and D.~Teaney,
Phys.\ Rev.\  C {\bf 81}, 034907 (2010).

\bibitem{Arnold:2008vd}
P.~B.~Arnold and W.~Xiao,
Phys.\ Rev.\  {\bf D78}, 125008 (2008).


\end{thebibliography}
\end{document}